\documentclass{scrartcl}

\usepackage[utf8]{inputenc}

\usepackage{amsbsy,amsfonts,amsmath,amssymb,enumerate,epsfig,graphicx,rotating,subfigure}

\usepackage{graphicx}
\usepackage{multicol}
\usepackage{minted} 

\newcommand{\Walberla}{\textsc{waLBerla}}

\renewcommand {\listingscaption}{\setlength{\leftskip}{1cm} \fontsize{8}{9}\selectfont Listing }

\newcounter{MyListingsCounter}
\newcommand{\mylistingscaption}[1]{\vspace{0.3cm} \fontsize{8}{9}\selectfont Listing \arabic{MyListingsCounter}.  #1}

\newcommand{\mylistingslabel}[1]{\addtocounter{MyListingsCounter}{-1}\refstepcounter{MyListingsCounter}\label{#1}\addtocounter{MyListingsCounter}{1}}

\include{pygmentsDefs}

\begin{document}

\title{A Python Extension for the Massively Parallel Multiphysics Simulation Framework waLBerla}


\author{\footnotesize Martin Bauer$^\dagger$, 
	Florian Schornbaum$^\dagger$, 
	Christian Godenschwager$^\dagger$, 
	Matthias Markl$^\dagger$,
	Daniela Anderl$^\dagger$,\\
	{\footnotesize 
	Harald Köstler$^\dagger$,
	and Ulrich Rüde$^\dagger$ } \\
    \vspace{0.3cm}
    {\footnotesize  $\dagger$ 
  		       Friedrich-Alexander-Universit\"at Erlangen-N\"urnberg, Cauerstra\ss e 11, Erlangen, Germany }\\
  }

\date{ }

\maketitle


\begin{abstract}

We present a Python extension to the massively parallel HPC simulation toolkit \Walberla{}.
\Walberla{} is a framework for stencil based algorithms operating on block-structured grids, 
with the main application field being fluid simulations in complex geometries using the lattice Boltzmann method.
Careful performance engineering results in excellent node performance and good scalability to over 400,000 cores.
To increase the usability and flexibility of the framework, a Python interface was developed.
Python extensions are used at all stages of the simulation pipeline: They simplify and automate scenario setup, evaluation, and plotting. 
We show how our Python interface outperforms the existing text-file-based configuration mechanism, providing features like automatic
nondimensionalization of physical quantities and handling of complex parameter dependencies. Furthermore, Python is used to 
process and evaluate results while the simulation is running, leading to smaller output files and the possibility to adjust 
parameters dependent on the current simulation state. 
C++ data structures are exported such that a seamless interfacing to other numerical Python libraries is possible.
The expressive power of Python and the performance of C++ make development of efficient code with low time effort possible.

\end{abstract}

\section{Introduction}

Many massively parallel codes are written for a specific use-case, making strict
assumptions on the scenario being simulated. These restrictions allow the programmer 
to optimize the code for its specific use-case, exploiting information already available at compile time. 
This approach is not feasible when developing a general purpose framework targeted at a variety of different applications.
While the highest priority is still performance and scalability, at the same time the framework has to be
easy to use, modular, and extensible \cite{dongarra2007impact}. However, performance and flexibility requirements are not necessarily conflicting goals. 

A common approach is to separate compute intensive parts of the program, so-called kernels,
and write several versions of them, each one being optimized for a special scenario or a specific target architecture.
The framework then selects the kernel which matches the problem and the hardware best.
Kernels typically are developed in low-level programming languages like C/C++ or Fortran that allow close control over the hardware.
These system programming languages are very powerful but also difficult to learn. The complex and subtle rules prevent possible
library users with a background in engineering who ofen only have a limited programming expertise to use parallel high performance codes. 
While being the best choice for performance critical portions of the code, these languages are therefore not well suited for other less time critical framework parts, like
simulation setup, simulation control, and result evaluation. The run time of these management tasks is usually negligible compared to kernel run times, since they
do not have to be executed as often as the compute kernels or are per-se less compute intensive.
Therefore, these routines are especially suitable for implementing them in a higher-level language like Python.

We present a Python extension to the massively parallel HPC framework \Walberla{} that aims to increase the ease of use 
and decrease the development time of non time critical functions.

\section{Related Work}

Python is a popular language in scientific computing for providing high level interfaces to fast subroutines and kernels
written in hardware near programming languages like Fortran or C/C++.
Numerous software packages exist that follow this wrapping approach: One example is the Trilinos multiphysics framework~\cite{heroux2005overview} providing 
wrappers for their packages in the PyTrilinos subproject~\cite{PyTrilinos}. Similarly, the finite volume library Clawpack (``Conservation Laws Package'') 
also provides high level interfaces through a Python package called PyClaw~\cite{ketcheson2012pyclaw}. These interfaces can then be used to couple
different frameworks~\cite{alghamdi2011petclaw,mandli2011using}.

For the lattice Boltzmann method, examples of Python-enabled frameworks are the Palabos software~\cite{palabosWebsite} 
and the GPU-focused SailFish library~\cite{januszewski2014sailfish}.

Additionally other packages use Python for generating code in hardware near languages from domain-specific languages~\cite{terrel2011pythonCodeGeneration}. 
One prominent example for this approach is the FEniCS project~\cite{logg2012fenics} for the automated solution of differential equations using finite elements. 
It uses the form compiler FFC~\cite{kirby2006compiler} to translates a domain-specific language into C++ code. 
This promising technique is also used in \Walberla{} for generating code for various lattice
Boltzmann stencils (D3Q19, D3Q27, D2Q9, etc.) but will not be described further in this paper where we focus on the simplification of the simulation workflow
by providing a high level Python interface to the core C++ framework.


\section{waLBerla multiphysics framework }

\Walberla{} is a massively parallel software framework supporting a wide range of scientific applications. 
Its main application are simulations based on the lattice Boltzmann method (LBM), which is reflected in the acronym ``widely applicable lattice Boltzmann from Erlangen''.
Having initially been a framework for the LBM, \Walberla{} evolved over time into a general purpose HPC framework for algorithms that can make use of a block-structured
domain partitioning. It offers data structures for implementing stencil based algorithms together with
load balancing mechanisms and routines for efficient input and output of simulation data.
\Walberla{} has two primary design goals: Being efficient and scalable on current supercomputer architectures,
while at the same time being flexible and modular enough to support various applications \cite{Feichtinger2011105,walberlaWebsite}.

In this section, we start by giving a short overview of the lattice Boltzmann free surface method, followed by a description of the \Walberla{} software stack used to implement the method.

\subsection{Free Surface Lattice Boltzmann Method}

The lattice Boltzmann method is a mesoscopic method for solving computational fluid dynamics (CFD) problems. It is based on a discrete
version of the Boltzmann equation for gases. The continuous Boltzmann equation comes from kinetic theory and reads \cite{Aidun2010}:
\[ \frac{\partial f}{\partial t} + \xi \cdot \nabla f = Q(f,f),  \]
with $f(x,\xi,t)$ being the continuous probability density function representing the probability of meeting a particle with velocity $\xi$ at position $x$ at time $t$. 
The left hand side of the equation describes the transport of particles, whereas the right hand side $Q(f,f)$ stands for a general particle collision term. 

To discretize the velocity space, a D3Q19 stencil with 19 discrete velocities $\{ e_\alpha | \alpha=0,...,18\}$ and 
corresponding particle distribution functions (PDFs) denoted by $f_\alpha(x,t)$ is used~\cite{Qian1992} .
With a time step length of $\Delta t$, the discrete LB evolution equation then reads:
\[ f_\alpha(x_i+e_\alpha \Delta t,t+\Delta t) - f_\alpha(x_i,t) = \Omega_\alpha(f). \]
As discrete LBM collision operator $\Omega_\alpha(f)$, a two relaxation time scheme (TRT) is used~\cite{Ginzburg2008, Ginzburg2008a}.
The time and space discretization yield an explicit time stepping scheme on a regular grid, that enables efficient parallelization due to the strict locality of the scheme.
To update the $19$ PDFs stored in a cell, only PDFs from the cell itself and neighboring cells are required.

This basic LBM is extended to facilitate the simulation of two-phase flows. The free surface lattice Boltzmann method (FSLBM)
is based on the assumption that the simulated liquid-gas flow is completely dominated by the heavier liquid phase such that the dynamics of the
lighter gas phase can be neglected. The problem is reduced to a single-phase flow with a free boundary~\cite{korner2005lattice}.
Following a volume of fluid approach, for each cell a {\em fill level} $\varphi$ is stored, representing the volume fraction of the heavier fluid in each cell.
The fill level determines the state of a cell: cells entirely filled with heavier fluid ($\varphi=1$) are marked as a {\em liquid cell} and are simulated by the LBM described above, whereas in 
{\em gas cells} ($\varphi=0$) no treatment with the LBM is necessary.
Between the two phases, in cells where $0<\varphi<1$, a closed layer of so-called {\em interface cells} is maintained, tracking all cells where the free boundary condition has to be applied.
A mass advection algorithm modifies the fill level $\varphi$ and triggers conversions of cell states. 

Additionally, regions of connected gas cells are tracked with a special {\em bubble model}~\cite{donath2009localized}.
This model calculates the volume of bubbles in order to compute the pressure for each gas cell as fraction of current to initial bubble volume.
The gas pressure is essential for the treatment of the free boundary.
Possible topology changes of bubbles require a sophisticated parallel algorithm to track bubble coalescence and break-up.

\subsection{Software Architecture}

\begin{figure}
 \begin{center}
    \subfigure[software architecture]{
      \resizebox*{8cm}{!}{  \includegraphics{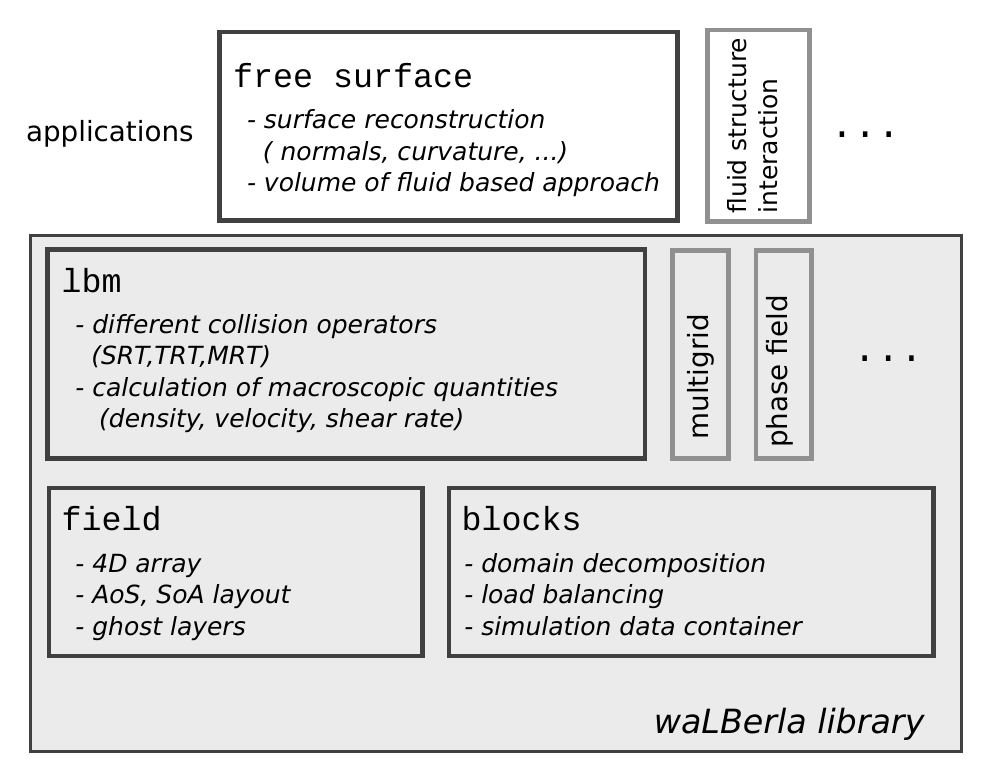}  } 
     }
     \hspace{0.8cm}
    \subfigure[domain partitioning]{
      \resizebox*{4cm}{!}{  \includegraphics{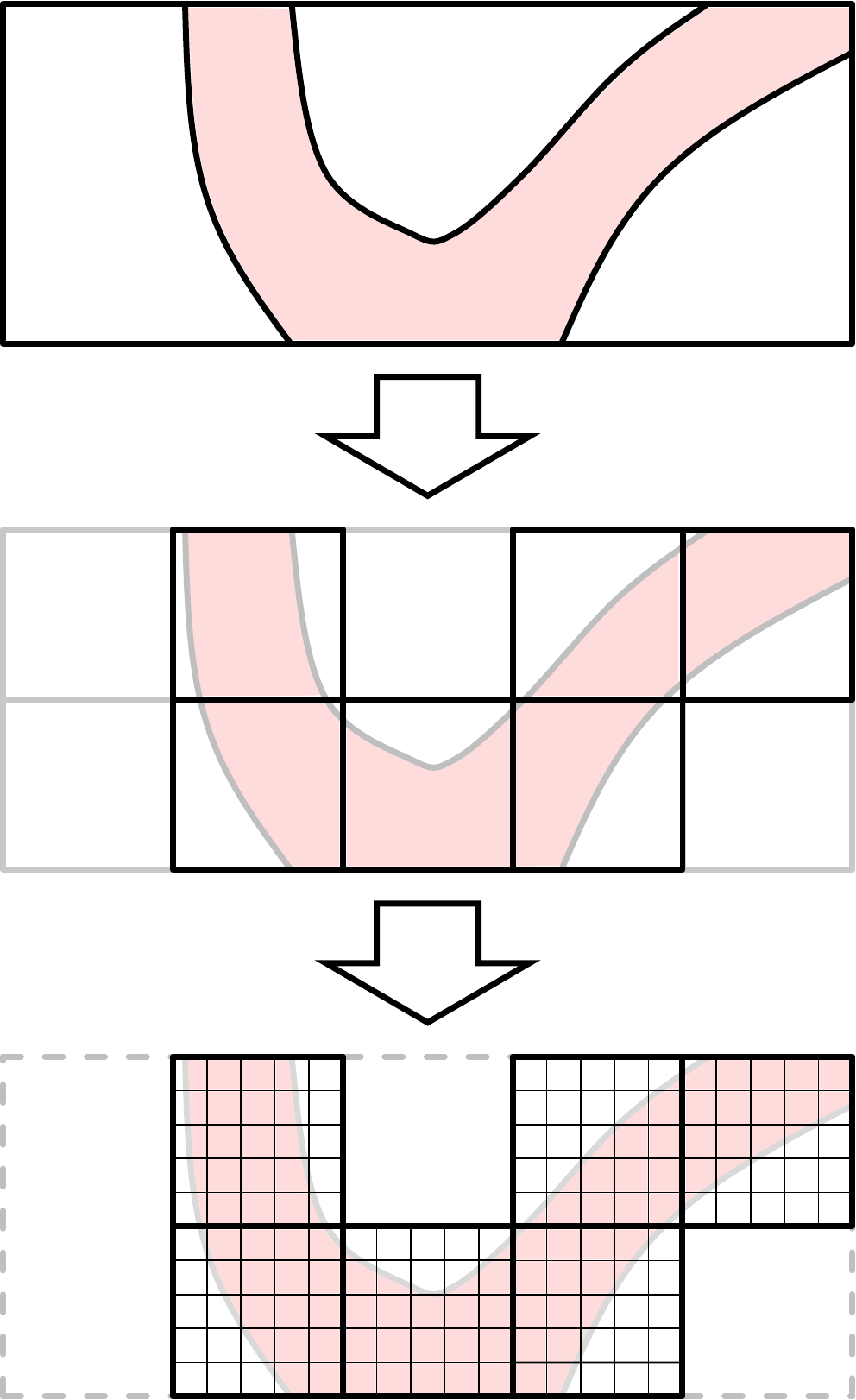}
     }
    }
     \caption{ Overview of waLBerla software architecture (left) and illustration of domain partitioning (right)}
     \label{fig:SoftwareArch_DomainPartitioning}
 \end{center}
\end{figure}

\Walberla{} is built out of a set of modules that can be grouped into three layers (Fig. \ref{fig:SoftwareArch_DomainPartitioning}a).
The bottom layer of \Walberla{} provides data structures and functions for implementing stencil-based algorithms on block-structured grids and will be described in
detail in this section. The second layer consists of specific algorithms which make use of the core layer. \Walberla{} is predominantly used for lattice Boltzmann simulations,
but also phase field and multigrid methods for finite differences and finite volumes have been implemented using the framework. 
The topmost layer is formed by algorithm extensions, like methods for fluid structure interaction \cite{Goetz2008,markl2015numerical} or the free surface LBM studied in this paper.

\subsubsection{Domain Decomposition}

For parallel simulations, the domain is partitioned into smaller, equally sized sub-domains called {\em blocks} which are then
distributed to processes. Blocks are not only the basis for parallelization, but also the basic unit of 
load balancing. Since there might be different computational efforts required to process each block, it is possible to put more
than one block on a process, balancing the computation time across all processes~\cite{schornbaum2015massively}.

To illustrate this concept, consider a simulation scenario as depicted in Figure~\ref{fig:SoftwareArch_DomainPartitioning}b.
In this scenario, the blood flow through an artery tree, which is specified by a triangle surface mesh, has to be simulated.
As a first step, the bounding box of the artery tree is decomposed into blocks, then blocks which do not overlap the mesh are discarded.
In the second step, the blocks are assigned to processes, taking the computational load and memory requirements of a block into account. In this example,
the computational load of a block is proportional to the number of fluid cells contained in it. 
The load balancing also takes into account neighborhood relations of blocks and the amount of data which has to be communicated between processes.

In a second step, a Cartesian grid (field) is allocated on each block where numerical kernels can be implemented efficiently due to its simple array structure. 
This two stage approach of partitioning the domain into blocks using octrees and later storing a Cartesian grid on them provides a good balance
between flexibility and performance and was recently also adopted in a new software project called ForestClaw~\cite{Burstedde2014forestclaw}.

\subsubsection{Fields}

Besides being the basic unit of load balancing, blocks also act as containers for distributed simulation data structures.
In the case of simulations with the LBM, the main data structure is the lattice. 
This lattice is fully distributed to all blocks, where the local part of the lattice is represented by an instance of the {\em field} class (last stage of Fig. 2).

Fields are implemented as four dimensional arrays: three dimensions for space and one dimension to store multiple values per cell.
In the LBM case, this fourth coordinate is used to store the 19 PDF values.
The field abstraction makes it possible to switch between an array-of-structures (AoS) and a structure-of-arrays (SoA) memory layout easily.
For many stencil algorithms, a AoS layout is beneficial since in this case all values of a cell are stored
consecutively in memory. This data locality results in an efficient usage of caches.
However, when optimizing algorithms to make use of SIMD instruction set extensions,
usually a SoA layout is better suited. Additionally, operands of SIMD instructions have to be 
aligned in memory, resulting in the requirement that the first elements of each line
are stored at aligned memory locations. 
To fulfill this restriction, additional space has to be allocated at the end of a cache line (padding).
This is implemented in the field class by discriminating between the requested size of a coordinate and the
allocated size for this coordinate. This discrimination is also helpful when implementing sliced views on fields, 
which operate on the original field data, but have different sizes.

\Walberla{} offers a synchronization mechanism for fields based on ghost layers. The field is extended by
one or more layers to synchronize cell data on the boundary between neighboring blocks. The neighbor access pattern
of the stencil algorithm determines the number of required ghost layers: If only next neighbors are accessed as in the LBM case,
one ghost layer is sufficient. Accessing cells further away requires more ghost layers.

\subsection{Performance and Scalability}

\begin{figure}
 \begin{center}
      \resizebox*{10cm}{!}{  \includegraphics{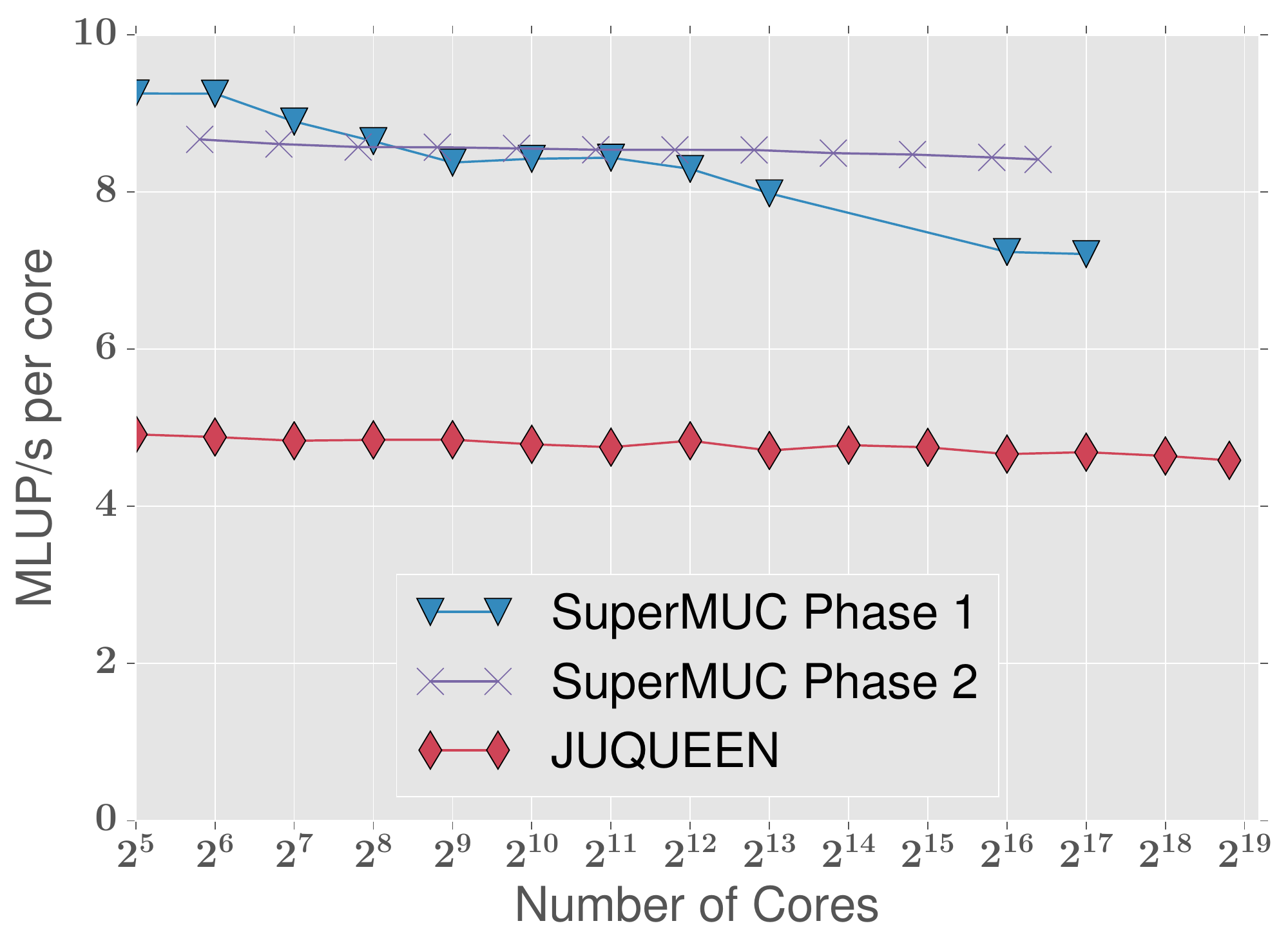}  } 
     
     \caption{ LBM weak scaling results with \Walberla{} on the top supercomputers in Germany: SuperMUC and JUQUEEN  }
     \label{fig:LBMScaling}
 \end{center}
\end{figure}

The \Walberla{} framework has been run on a wide range of supercomputing clusters, for example on \textsc{JUQUEEN} in J\"{u}lich, 
on Tsubame at the GSIC Center at the Tokyo Institute of 
Technology in Japan, and on \textsc{SuperMUC} at LRZ in Munich \cite{Goetz:2010:ParComp,feichtinger2015performance}. 

Figure \ref{fig:LBMScaling} illustrates weak scaling results obtained when running a LBM-based fluid simulation on a dense regular domain 
with \Walberla{} on SuperMUC in Munich and the JUQUEEN system in Jülich. On SuperMUC, scaling results are shown for the Phase 1 system installed in 2012
as well as the newer Phase 2 system from 2015. Performance is measured in million lattice site updates per second (MLUP/s) per core, i.e., how many lattice Boltzmann cells are updated 
by one core in one second.
The scaling experiments show that \Walberla{} can make use of the full machines with a high parallel efficiency.
The largest simulation run on JUQUEEN resolved the simulation domain with over a trillion cells. 
Further weak and strong scaling results of \Walberla{} can be found in \cite{godenschwager2013,LSS:Bauer:2015}.


\section{Python Interface}

This section gives an overview of the Python interface to the basic \Walberla{} data structures described in the previous section. 

Initially, \Walberla{} was designed as a pure C++ framework. The Python interface is developed as an optional extension to the framework.
The motivation for using Python originated from the need for a more flexible simulation setup mechanism. A text file was used to configure the simulation, which became
increasingly complex over time. Some users wrote Python scripts to create this configuration file, leading to the idea to embed Python directly into \Walberla{}.
It turned out that the embedding of Python is useful, not only for configuration purposes, but also for simulation control and analysis as described in the following section.

The C++ part of waLBerla makes use of various {\em boost} libraries~\cite{boostWebsite}, for example for portable filesystem access, for memory management with smart pointers, and for parsing of
input parameters using regular expression. Since \Walberla{} already depends on {\em boost}, we also make use of the {\em boost::python} library~\cite{boostPythonWebsite}
to expose our C++ data structures to Python. For certain tasks, however, it was necessary to use the Python C-API directly, since the required functionality
is not available in {\em boost::python}.

There are two mechanisms for coupling C++ and Python. The first approach is to create a Python module as shared library out of the C++ code. 
Using this solution, the driving code is written in Python, making use of exposed C++ functionality in the library.
In the second approach, Python is embedded into the C++ application by linking
against {\em libpython}. When the second approach is used, the simulation is driven by C++ code, optionally calling Python functions at certain stages of the simulation.
We choose the second approach, since our main goal is to extend our C++ simulation code making it more flexible and easier to use.
However, additionally an implementation of the first approach is currently in development, offering the possibility to drive simulations using Python code~\cite{walberlaWebsite}.
An interactive demonstration of this first approach can be found at \cite{walberlaWebsite}.

To interact with C++ simulation code via Python, the user supplies a script file decorated with callback functions as shown in Listing \ref{lst:embedding_py}.
The code example shows a callback function as it is often used for custom post-processing or monitoring of the current simulation.

\begin{listing}[H]

\begin{minipage}[c]{0.45\textwidth}
\vspace{0.2cm}
\begin{Verbatim}[commandchars=\\\{\},fontsize=\footnotesize]
\PY{k+kn}{import} \PY{n+nn}{waLBerla}

\PY{n+nd}{@waLBerla.callback}\PY{p}{(} \PY{l+s}{\PYZdq{}}\PY{l+s}{at\PYZus{}end\PYZus{}of\PYZus{}timestep}\PY{l+s}{\PYZdq{}} \PY{p}{)}
\PY{k}{def} \PY{n+nf}{my\PYZus{}callback}\PY{p}{(} \PY{n}{blockstorage}\PY{p}{,} \PY{o}{*}\PY{o}{*}\PY{n}{kwargs} \PY{p}{)}\PY{p}{:}
  \PY{k}{for} \PY{n}{block} \PY{o+ow}{in} \PY{n}{blockstorage}\PY{p}{:}
    \PY{c}{\PYZsh{} access local simulation data}
    \PY{n}{velocity\PYZus{}field} \PY{o}{=} \PY{n}{block}\PY{p}{[}\PY{l+s}{\PYZsq{}}\PY{l+s}{velocity}\PY{l+s}{\PYZsq{}}\PY{p}{]} 
\end{Verbatim}
\vspace{0.2cm}
\end{minipage}
\hspace{0.02\textwidth}\vrule\hspace{0.02\textwidth}
\begin{minipage}[c]{0.45\textwidth}
\vspace{0.2cm}
\begin{Verbatim}[commandchars=\\\{\},fontsize=\footnotesize]
\PY{n}{Callback} \PY{n}{cb} \PY{p}{(} \PY{l+s}{\PYZdq{}}\PY{l+s}{at\PYZus{}end\PYZus{}of\PYZus{}timestep}\PY{l+s}{\PYZdq{}} \PY{p}{)}\PY{p}{;}
\PY{n}{cb}\PY{o}{.}\PY{n}{exposePtr}\PY{p}{(}\PY{l+s}{\PYZdq{}}\PY{l+s}{blockstorage}\PY{l+s}{\PYZdq{}}\PY{p}{,} \PY{n}{blockStorage} \PY{p}{)}\PY{p}{;}
\PY{n}{cb}\PY{p}{(}\PY{p}{)}\PY{p}{;} \PY{o}{/}\PY{o}{/} \PY{n}{run} \PY{n}{python} \PY{n}{function}
\end{Verbatim}
\vspace{0.2cm}
\end{minipage}

\mylistingscaption{Embedding Python into C++ using annotations. C++ code (right) calls Python code (left).}
\mylistingslabel{lst:embedding_py}

\end{listing}

In this case, the function is called after a simulation time step has finished, such that all data is in a consistent state. 
The callback mechanism exposes all simulation data, passing the blockstorage object to the function.
For simple and intuitive access, the block collection is exposed as a mapping type, mimicking the behavior
of a Python dictionary.
In parallel simulations, the Python callback function is invoked on every process, whereas the blockstorage contains only blocks assigned to the current process.
Thus, the loop over all blocks is implicitly parallelized. 
If a global quantity has to be calculated, the data reduction has to be programmed manually using MPI routines.

The C++ counterpart of the callback function is shown in Listing \ref{lst:embedding_py} on the right. A callback object is created, identified by a string which has to match the decorator 
string in the Python script. Then the function arguments are passed, either by reference (``exposePtr``) or by value (''exposeCopy``).
In order to pass a C++ object to the callback, the class has to be registered for export using mechanisms provided by {\em boost::python }.
For most data structures, this can be done in a straightforward way, for the field class however, a special approach has to be taken.

\subsection{waLBerla Field as NumPy Array}

The field class is one of the central data structures of \Walberla{}. It is the object that programmers have to work with the most, for example when
setting up simulation geometry and boundary conditions or when evaluating and analyzing the current simulation state. 
As described above, a field is essentially a four dimensional array supporting different memory layouts (AoS and SoA), aligned allocation strategies and advanced indexing (slicing).

A similar data structure widely used in the Python community is \texttt{ndarray} provided by the NumPy package~\cite{van2011numpy}.
A wide range of algorithms exist operating on NumPy arrays, for example linear algebra, Fourier transformation, or image processing routines.
To make use of these algorithms, it is desirable to be able to convert \Walberla{} fields efficiently into the NumPy representation.
Copying data between these representations is not a feasible option due to performance reasons and memory limitations. Simulations are oftentimes set up in a way to 
fully utilize the available memory of a compute node. Large portions of the allocated memory are occupied by the lattice, i.e., the field.
Thus, an export mechanism for fields is required, which offers read-write access to the field {\em without copying data}. 
The exposed object should behave like a NumPy array such that algorithms from the NumPy and SciPy ecosystem can be used.
There are only two options fulfilling these requirements: One possibility is to essentially re-implement \texttt{ndarray} and to export all functions using mechanisms provided by {\em boost::python}.
This is the {\em duck-typing} approach popular in Python: The exported field would behave exactly like a \texttt{ndarray} and could therefore be used with all algorithms expecting NumPy arrays.
Due to the high implementation effort, this approach was not used. Instead, we implement the Python {\em buffer protocol}~\cite{pythonBufferProtocol}
which provides a standardized way to directly access memory buffers. This protocol supports the advanced memory layouts used by the \Walberla{} field class through definition of strides and offsets.
Among others, NumPy arrays can be constructed from buffer objects, so all requirements can be fulfilled using this approach, without introducing any dependency of \Walberla{} to the NumPy library.
The buffer protocol is not available in {\em boost::python}, so in this case the C-API of Python had to be used directly.

\subsection{Encountered Difficulties}

The two primary goals of \Walberla{}, being an HPC framework, are flexibility and performance.
To achieve both goals, it is necessary to make use of advanced C++ template mechanisms.
Exporting these template constructs to Python can be difficult since all templates have to be instantiated with
all possible parameter combinations. In situations where this cannot be done manually, the instantiation is done
using template meta programming.
Instantiating and exporting all possible template parameter combinations would result in long compile times and would increase the size of 
the executable significantly. Thus, a tradeoff has to be made and only the commonly used template parameter combinations are exported to Python.
If, however, users need other combinations in their applications, the framework provides a simple mechanism to configure which template parameter combinations are exported.

	
\section{Simplification of Simulation Workflow} 

In this section, we demonstrate the usage of the \Walberla{} Python interface by 
describing the workflow of setting up two different FSLBM scenarios.
We illustrate how to configure, control, and evaluate the simulations using Python callback functions.

\begin{figure}
 \begin{center}
    \subfigure[]{
      \resizebox*{6cm}{!}{  \includegraphics{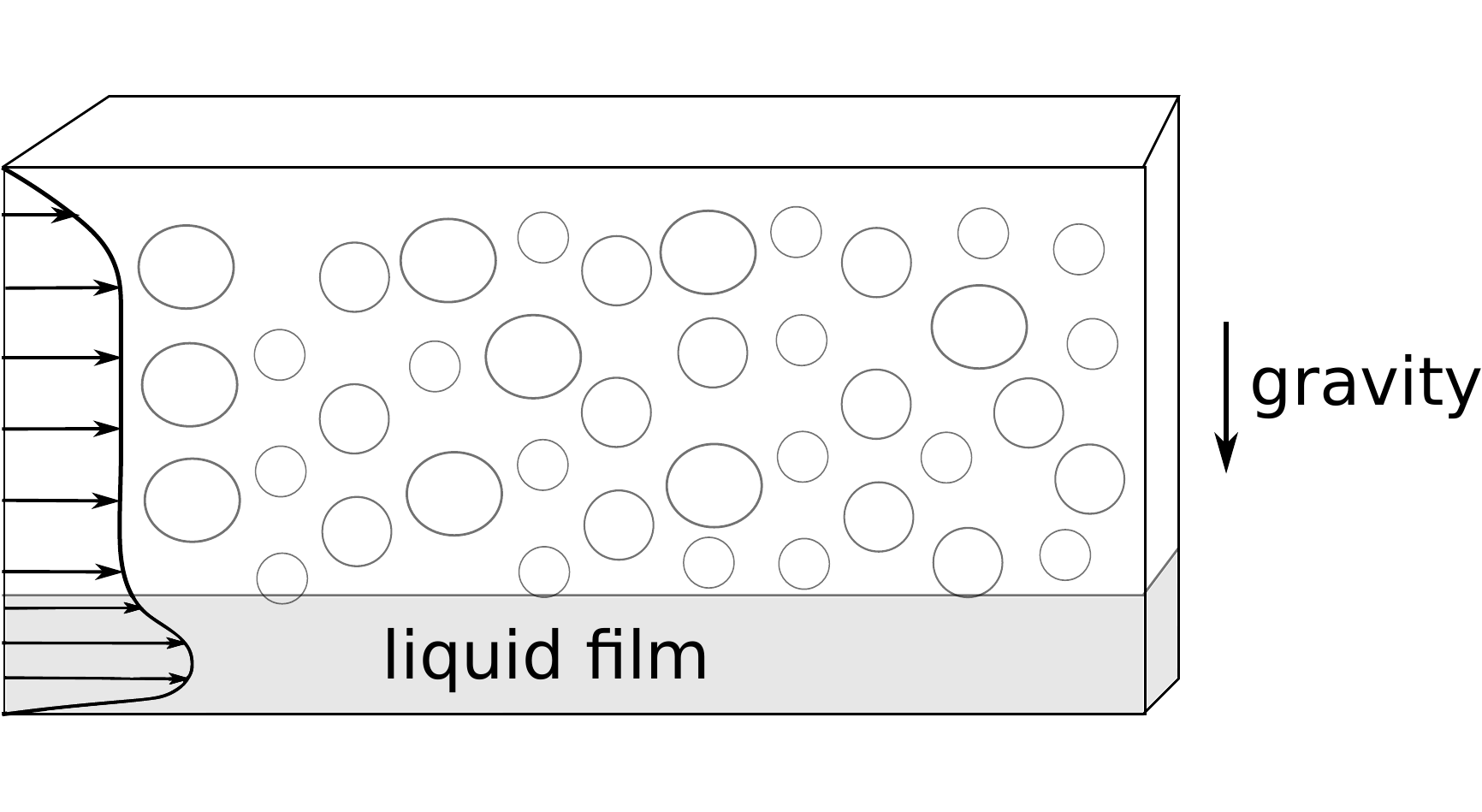}  } 
     }
     \hspace{0.8cm}
    \subfigure[]{
      \resizebox*{4cm}{!}{  \includegraphics{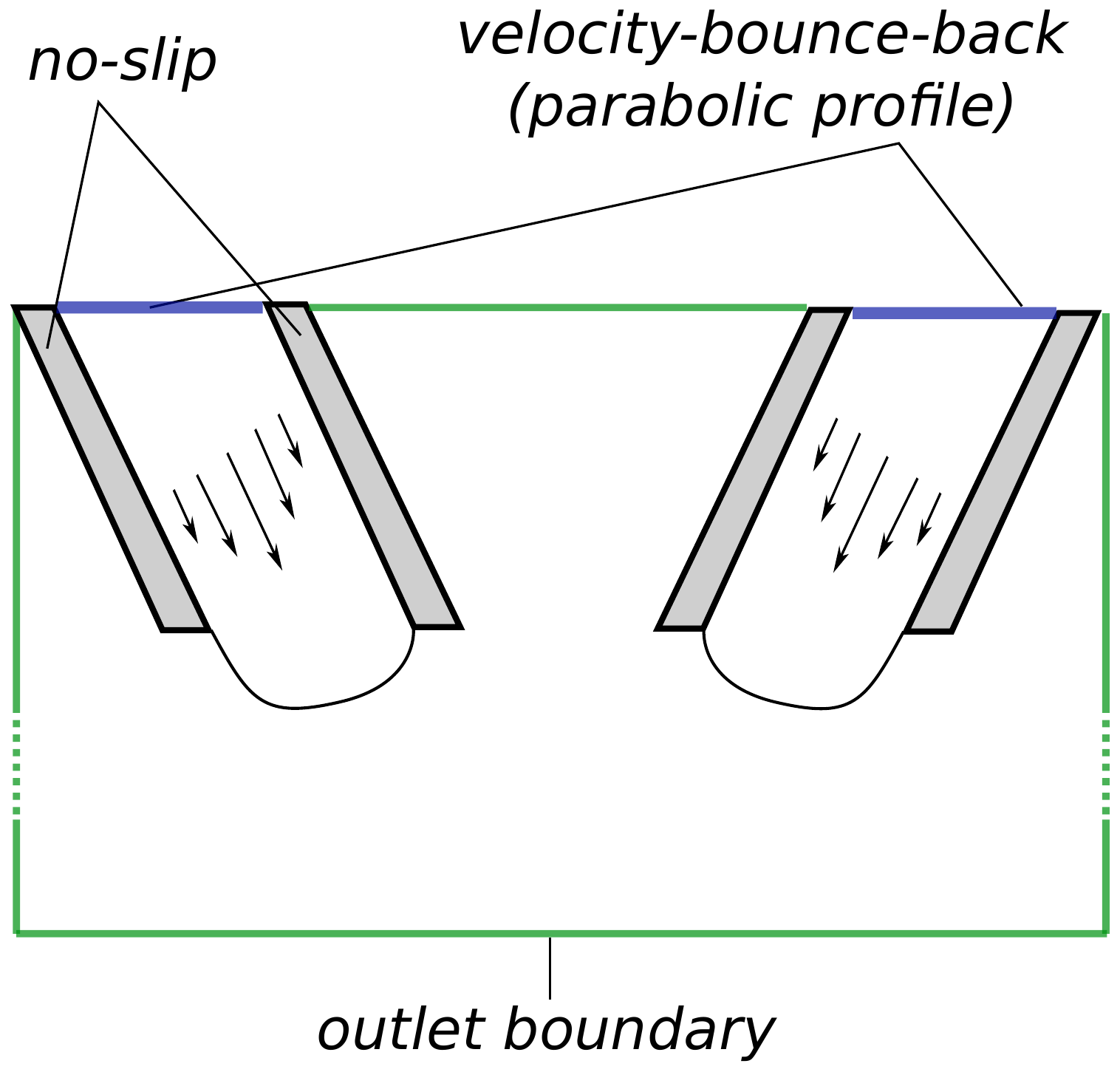} 
     }
    }
     \caption{ Simulation setup of bubbly flow through channel (a) and twin jet scenario (b) }
     \label{fig:scenario_setup}
 \end{center}
\end{figure}

The first example scenario is a two-phase flow problem in a rectangular channel.
A foam is transported through a channel as depicted in Figure~\ref{fig:scenario_setup}a.
Due to gravity, bubbles rise to the top of the channel, forming a liquid layer at the bottom. 
A characteristic flow profile is expected, with a parabolic shape in the liquid film, and an almost constant
velocity in the rest of the channel. The goal of the simulation is to investigate the stability of the transported foam and its dependence on surface properties and rheological parameters.
Desired output quantities of this simulation are gas fractions and velocities in different parts of the domain.
Additionally, the flow profile and foam stability should be evaluated when the pressure gradient which drives the channel is changed,
i.e., when the pressure gradient is switched on and off.

In the second example scenario shown in Figure~\ref{fig:scenario_setup}b, two inclined pipes generate jets which form a thin lamella when the jets hit each other (Figure~\ref{fig:jets_render_result}). 
The goal in this scenario is to determine the lamella size, thickness, and rim stability for different pipe geometries and rheological parameters. 
The simulation should be terminated automatically when the lamella is fully developed.

Using these two example scenarios, we describe how the Python interface simplifies the configuration and geometry setup of a simulation with the LBM.

\subsection{Simulation Setup}

To setup a simulation on a regular grid, there are typically two different kinds of input required. 
Before the Python interface was developed, the most flexible choice for domain initialization, geometry setup, and specification of boundaries was a voxel-based input file. 
Such voxel files had to be generated using external tools.
Additional configuration like discretization options or physical parameters where given in a second text file. This text file was formatted in a syntax similar to JavaScript object notation (JSON),
providing parameters as a hierarchy of key-value pairs.
When setting up LBM simulations, physical parameters have to be converted to nondimensionalized lattice units~\cite{junk2006}.
The conversion factors depend on the choice of discretization parameters. To simplify this process, the configuration file
was extended with special functionality, enabling the user to easily convert physical to lattice units.
The nondimensionalization problem is, however, only an instance of a wider range of problems. The problem that parameters are interdependent and 
that this interdependency should be defined by the user in the configuration file. Parameters can depend on each other in complex ways:
consider the time step length, which should be chosen maximal, subject to stability constraints.
Complex parameter dependencies can lead to configuration errors.
To improve usability, especially for inexperienced users, configuration mistakes should be detected
before the simulation runs. Therefore, all parameters have to be checked if they are in a valid range and consistent with other parameters.

Trying to handle these problems in a flexible and user friendly way leads towards more and more custom extensions in the input file, essentially developing
a custom scripting language. Instead of pursuing this approach further, the decision was made to use an existing scripting language like Python.
Python offers libraries that can handle all of the requirements described above. There are libraries available for defining and manipulating physical units, 
making them suitable for solving nondimensionalization problems~\cite{pintWebsite}.
To handle complex parameter dependencies, we use linear algebra and optimization routines from {\em SciPy}~\cite{jones2001scipy}.
Also, the ability for symbolic calculations as provided by the {\em SymPy} library~\cite{joyner2012sympy} proves useful in a configuration file.

The hierarchical key-value configuration is represented in Python as a nested dictionary object. This dictionary is built up in a specially decorated function called by
the C++ part of the framework (first function in Listing \ref{lst:simulation_setup_py}). 
For simplicity, the C++ part expects all parameters to be in valid nondimensionalized lattice units. Nondimensionalization and parameter validation is completely done in Python.

The definition of domain geometry and boundary conditions can also be handled using a Python callback function, substituting the previously used voxel file.
This callback is executed once for every cell before the simulation starts. Via the returned dictionary object, the initial cell state is defined, 
consisting of initial velocity and density, or of the boundary type and boundary parameters.

\begin{listing}[H]

\begin{minipage}[c]{0.48\textwidth}
\vspace{0.2cm}
\begin{Verbatim}[commandchars=\\\{\},fontsize=\footnotesize]
\PY{n+nd}{@waLBerla.callback}\PY{p}{(} \PY{l+s}{\PYZdq{}}\PY{l+s}{config}\PY{l+s}{\PYZdq{}} \PY{p}{)}
\PY{k}{def} \PY{n+nf}{config}\PY{p}{(}\PY{p}{)}\PY{p}{:}
  \PY{n}{c} \PY{o}{=} \PY{p}{\PYZob{}} 
    \PY{l+s}{\PYZsq{}}\PY{l+s}{Physical}\PY{l+s}{\PYZsq{}} \PY{p}{:} \PY{p}{\PYZob{}}
     \PY{l+s}{\PYZsq{}}\PY{l+s}{viscosity}\PY{l+s}{\PYZsq{}}      \PY{p}{:} \PY{l+m+mf}{1e\PYZhy{}6}\PY{o}{*}\PY{n}{m}\PY{o}{*}\PY{n}{m}\PY{o}{/}\PY{n}{s}\PY{p}{,}
     \PY{l+s}{\PYZsq{}}\PY{l+s}{surface\PYZus{}tension}\PY{l+s}{\PYZsq{}}\PY{p}{:} \PY{l+m+mf}{0.072}\PY{o}{*}\PY{n}{N}\PY{o}{/}\PY{n}{m}
     \PY{l+s}{\PYZsq{}}\PY{l+s}{dx}\PY{l+s}{\PYZsq{}}             \PY{p}{:} \PY{l+m+mf}{0.01}\PY{o}{*}\PY{n}{m}\PY{p}{,}
    \PY{p}{\PYZcb{}}
    \PY{l+s}{\PYZsq{}}\PY{l+s}{Control}\PY{l+s}{\PYZsq{}} \PY{p}{:} \PY{p}{\PYZob{}}
     \PY{l+s}{\PYZsq{}}\PY{l+s}{timesteps}\PY{l+s}{\PYZsq{}} \PY{p}{:} \PY{l+m+mi}{10000}\PY{p}{,}
     \PY{l+s}{\PYZsq{}}\PY{l+s}{vtk\PYZus{}output\PYZus{}interval}\PY{l+s}{\PYZsq{}}\PY{p}{:} \PY{l+m+mi}{100}\PY{p}{,}
    \PY{p}{\PYZcb{}}
  \PY{p}{\PYZcb{}}
  \PY{n}{compute\PYZus{}derived\PYZus{}parameters}\PY{p}{(}\PY{n}{c}\PY{p}{)}
  \PY{n}{c}\PY{p}{[}\PY{l+s}{\PYZsq{}}\PY{l+s}{Physical}\PY{l+s}{\PYZsq{}}\PY{p}{]}\PY{p}{[}\PY{l+s}{\PYZsq{}}\PY{l+s}{dt}\PY{l+s}{\PYZsq{}}\PY{p}{]} \PY{o}{=} \PY{n}{find\PYZus{}optimal\PYZus{}dt}\PY{p}{(}\PY{n}{c}\PY{p}{)}
  \PY{n}{nondimensionalize}\PY{p}{(}\PY{n}{c}\PY{p}{)}  
  \PY{k}{return} \PY{n}{c}
\end{Verbatim}
\vspace{0.2cm}
\end{minipage}
\hspace{0.02\textwidth}\vrule\hspace{0.02\textwidth}
\begin{minipage}[c]{0.48\textwidth}
\vspace{0.2cm}
\begin{Verbatim}[commandchars=\\\{\},fontsize=\footnotesize]
\PY{n}{gas\PYZus{}bubbles}\PY{o}{=}\PY{n}{sphere\PYZus{}pack}\PY{p}{(}\PY{l+m+mi}{300}\PY{p}{,}\PY{l+m+mi}{100}\PY{p}{,}\PY{l+m+mi}{100}\PY{p}{)}

\PY{n+nd}{@waLBerla.callback}\PY{p}{(} \PY{l+s}{\PYZdq{}}\PY{l+s}{domain\PYZus{}init}\PY{l+s}{\PYZdq{}} \PY{p}{)}
\PY{k}{def} \PY{n+nf}{boundary\PYZus{}setup}\PY{p}{(} \PY{n}{cell} \PY{p}{)}\PY{p}{:}
  \PY{n}{p\PYZus{}w} \PY{o}{=} \PY{n}{c}\PY{p}{[}\PY{l+s}{\PYZsq{}}\PY{l+s}{Physics}\PY{l+s}{\PYZsq{}}\PY{p}{]}\PY{p}{[}\PY{l+s}{\PYZsq{}}\PY{l+s}{pressure\PYZus{}W}\PY{l+s}{\PYZsq{}}\PY{p}{]}
  \PY{n}{boundary}\PY{o}{=}\PY{p}{[}\PY{p}{]}
  \PY{k}{if}   \PY{n}{is\PYZus{}at\PYZus{}border}\PY{p}{(} \PY{n}{cell}\PY{p}{,} \PY{l+s}{\PYZsq{}}\PY{l+s}{W}\PY{l+s}{\PYZsq{}}   \PY{p}{)}\PY{p}{:} 
    \PY{n}{boundary} \PY{o}{=} \PY{p}{[} \PY{l+s}{\PYZsq{}}\PY{l+s}{pressure}\PY{l+s}{\PYZsq{}}\PY{p}{,} \PY{n}{p\PYZus{}w} \PY{p}{]}
  \PY{k}{elif} \PY{n}{is\PYZus{}at\PYZus{}border}\PY{p}{(} \PY{n}{cell}\PY{p}{,} \PY{l+s}{\PYZsq{}}\PY{l+s}{E}\PY{l+s}{\PYZsq{}}   \PY{p}{)}\PY{p}{:} 
    \PY{n}{boundary} \PY{o}{=} \PY{p}{[} \PY{l+s}{\PYZsq{}}\PY{l+s}{pressure}\PY{l+s}{\PYZsq{}}\PY{p}{,} \PY{l+m+mf}{1.0} \PY{p}{]}
  \PY{k}{elif} \PY{n}{is\PYZus{}at\PYZus{}border}\PY{p}{(} \PY{n}{cell}\PY{p}{,} \PY{l+s}{\PYZsq{}}\PY{l+s}{NSTB}\PY{l+s}{\PYZsq{}}\PY{p}{)}\PY{p}{:} 
    \PY{n}{boundary} \PY{o}{=} \PY{p}{[} \PY{l+s}{\PYZsq{}}\PY{l+s}{noslip}\PY{l+s}{\PYZsq{}} \PY{p}{]}
  \PY{n}{fl} \PY{o}{=} \PY{l+m+mi}{1}\PY{o}{\PYZhy{}}\PY{n}{gas\PYZus{}bubbles}\PY{o}{.}\PY{n}{overlap}\PY{p}{(}\PY{n}{cell}\PY{p}{)}
  \PY{k}{return}\PY{p}{\PYZob{}}\PY{l+s}{\PYZsq{}}\PY{l+s}{fill\PYZus{}level}\PY{l+s}{\PYZsq{}}\PY{p}{:}\PY{n}{fl}\PY{p}{,}
         \PY{l+s}{\PYZsq{}}\PY{l+s}{boundary}\PY{l+s}{\PYZsq{}}  \PY{p}{:}\PY{n}{boundary} \PY{p}{\PYZcb{}}
\end{Verbatim}
\vspace{0.2cm}
\end{minipage}

\mylistingscaption{Simulation setup of channel scenario. Extended key-value configuration (left) and simplified boundary setup (right). }
\mylistingslabel{lst:simulation_setup_py}

\end{listing}

\begin{figure}
\centerline{
   \includegraphics[width=0.8\columnwidth]{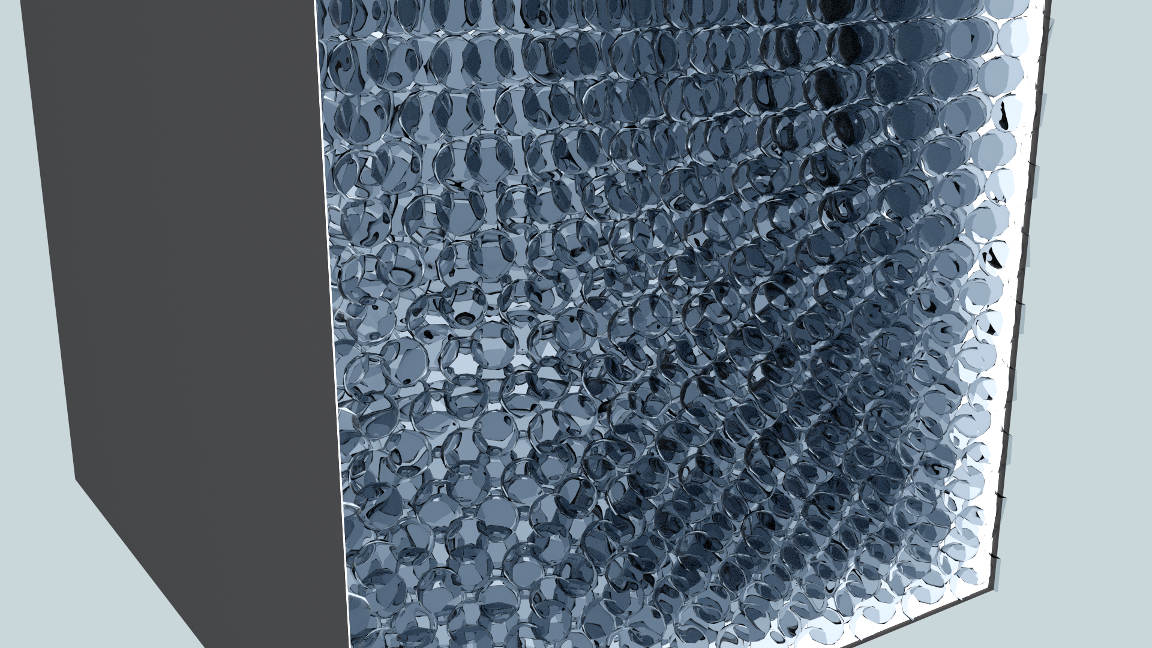}
}
\caption{Visualization of simulation result of first example scenario: Cross section orthogonal to flow direction}
\label{fig:channel_render_result}
\end{figure}

Listing \ref{lst:simulation_setup_py} shows how to set up the channel flow scenario using a Python file.
The first callback function substitutes the JSON file, providing parameters as a dictionary. 
Before passing the parameters to the C++ code, several functions operate on the dictionary, handling nondimensionalization and calculation of dependent parameters.

The second callback function handles boundary setup and domain initialization.
In our example scenario, we prescribe a pressure boundary on the left (east) and right (west) end of the domain, all other borders are set to no-slip boundary conditions.
In this example, only boundary conditions at domain borders are set. However, it is possible to set boundaries at arbitrary cells in the domain.
The bubbles are placed in the channel as a dense sphere packing, where bubble positions are calculated by a Python function. This routine fills the whole domain with 
equally sized bubbles. As shown in Listing \ref{lst:simulation_setup_py}, the initial gas fraction of a cell is set using the initialization callback mechanism.

Python is well suited for setting up more geometrically complex scenarios: For the impinging jets simulation, two rotated pipes have to be placed inside the domain
and a parabolic velocity profile has to be initialized inside each pipe. Therefor, a \texttt{Pipe} class is written using {\em numpy} for the $4x4$ transformation matrices which 
describe the mapping from world to object coordinates. In Python, this requires litte development effort and can be expressed in about 20 lines of code. 
Once the \texttt{Pipe} class is available, the simulation setup is straightforward (Listing \ref{lst:simulation_setup_impingingJets_py}).

\begin{listing}[H]
\begin{Verbatim}[commandchars=\\\{\},fontsize=\footnotesize]
\PY{n}{leftPipe} \PY{o}{=} \PY{n}{Pipe}\PY{p}{(} \PY{n}{diameter}\PY{p}{,} \PY{n}{length} \PY{p}{,} \PY{n}{leftPipePosition} \PY{p}{)}
\PY{n}{leftPipe}\PY{o}{.}\PY{n}{rotate}\PY{p}{(} \PY{l+m+mi}{45} \PY{p}{)}
\PY{c}{\PYZsh{} (...) same for rightPipe}

\PY{n+nd}{@waLBerla.callback}\PY{p}{(} \PY{l+s}{\PYZdq{}}\PY{l+s}{domain\PYZus{}init}\PY{l+s}{\PYZdq{}} \PY{p}{)}
\PY{k}{def} \PY{n+nf}{pipeSetup}\PY{p}{(} \PY{n}{cell} \PY{p}{)}\PY{p}{:}
  \PY{k}{for} \PY{n}{p} \PY{o+ow}{in} \PY{p}{[}\PY{n}{leftPipe}\PY{p}{,} \PY{n}{rightPipe}\PY{p}{]}\PY{p}{:}
    \PY{k}{if} \PY{n}{p}\PY{o}{.}\PY{n}{contains}\PY{p}{(}\PY{n}{cell}\PY{p}{)}\PY{p}{:}
      \PY{k}{return} \PY{p}{\PYZob{}} \PY{l+s}{\PYZsq{}}\PY{l+s}{initVel}\PY{l+s}{\PYZsq{}} \PY{p}{:} \PY{n}{p}\PY{o}{.}\PY{n}{parabolicVel}\PY{p}{(}\PY{n}{cell}\PY{p}{,}\PY{n}{maxVel}\PY{p}{)} \PY{p}{\PYZcb{}}
    \PY{k}{if} \PY{n}{p}\PY{o}{.}\PY{n}{shellContains}\PY{p}{(}\PY{n}{cell}\PY{p}{)}\PY{p}{:}
      \PY{k}{return} \PY{p}{\PYZob{}} \PY{l+s}{\PYZsq{}}\PY{l+s}{boundary}\PY{l+s}{\PYZsq{}} \PY{p}{:} \PY{l+s}{\PYZsq{}}\PY{l+s}{NoSlip}\PY{l+s}{\PYZsq{}} \PY{p}{\PYZcb{}} 
\end{Verbatim}

 \mylistingscaption{Boundary definitions for impinging jets scenario. }
 \mylistingslabel{lst:simulation_setup_impingingJets_py}
\end{listing}

\subsection{Evaluation}

Storing the complete state of a big parallel LBM simulation results in output files with sizes up to several terabytes.
Typically, the complete flow field together with cell fill levels is written to a voxel based file for analysis.
Especially for free surface simulations, not all of this detailed output is required. When simulating the behavior of foams,
only some higher level information like gas fractions in certain areas or number, shape, and velocity of bubbles are of interest.
These quantities can be obtained by a post-processing step using the raw voxel based output of velocity and fill level.
For moderately sized simulations, the raw output can be copied to a desktop machine and post-processed using graphical tools like ParaView~\cite{henderson2004paraview}.
For larger simulations, however, it has many advantages to do the post-processing ''in-situ`` directly on the cluster where the simulation is run.
The time for copying the raw output files can be saved and the post-processing algorithm itself can be parallelized if necessary.
Since the requirements of this evaluation step vary widely depending on the scenario at hand, the analysis routines are not included in the core C++ part of the framework.
The post-processing is usually done in user written, custom Python scripts which have a lower development time than C++ code.
This situation was another motivation to directly couple our C++ simulation framework to Python, such that evaluation scripts can run during the simulation and operate directly on
simulation data making the output of raw voxel data obsolete.

\begin{figure}
\centerline{
   \includegraphics[width=0.8\columnwidth]{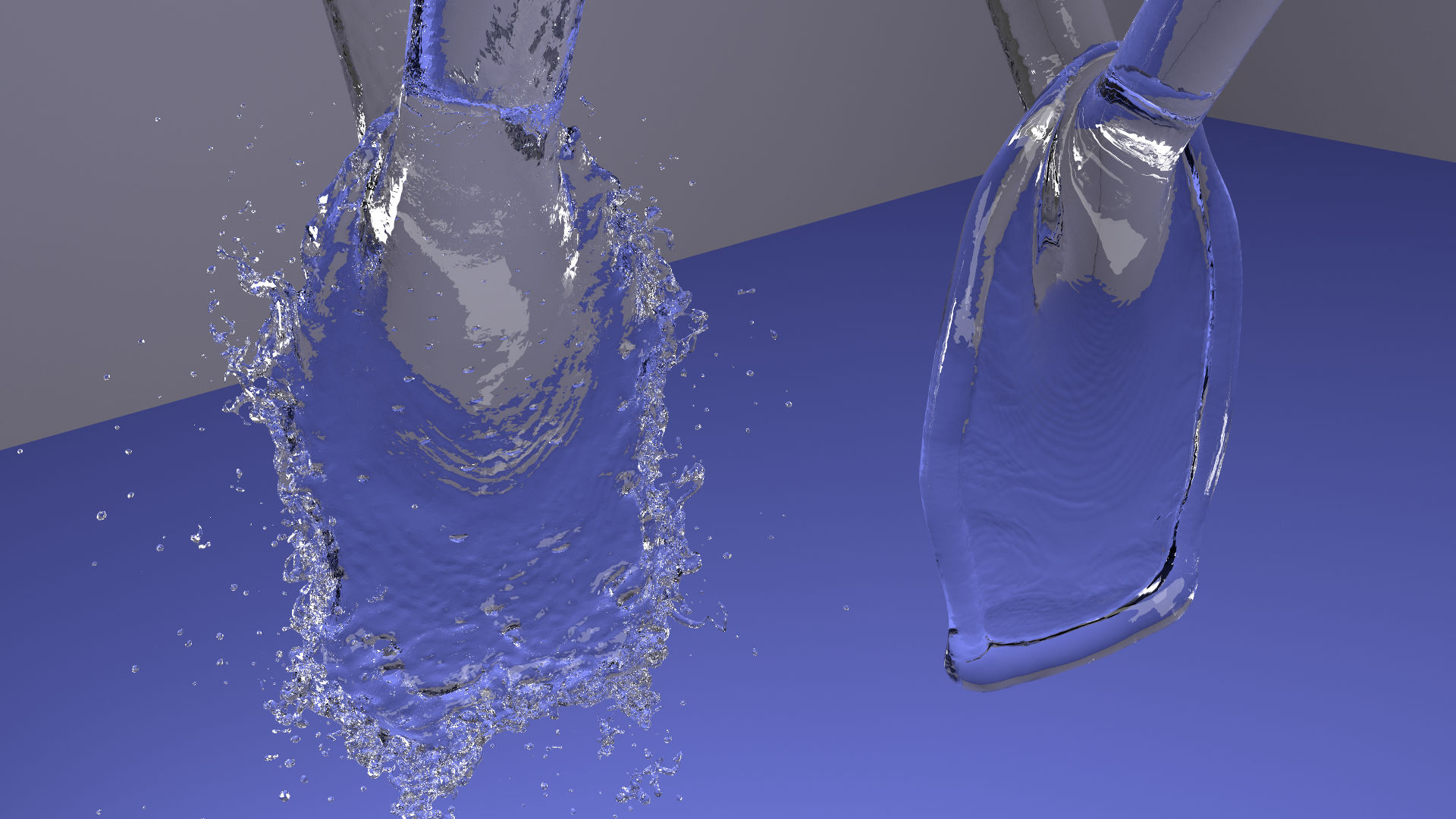}
}
\caption{Second example scenario: Impining jets simulated without (left) and with surface tension (right) }
\label{fig:jets_render_result}
\end{figure}

In the evaluation callback functions, we can make use of the exposed C++ data structures.
Since the simulation data is fully distributed, also the evaluation has to be done in a distributed way.
The evaluation code of the channel example scenario is shown in Listing \ref{lst:simulation_eval_py} where the maximum velocity along the flow direction
for the channel example problem is calculated.
The first iteration iterates all local blocks, extracting the velocity field as a NumPy array. 
As described above, the NumPy array is only a view on the already existing data, no copy is made. 
Using the velocity field, first the per-process maximum is determined, then a global MPI reduce operation has to be done
to obtain the global maximum.

\begin{listing}[H]
\begin{Verbatim}[commandchars=\\\{\},fontsize=\footnotesize]
\PY{k+kn}{import} \PY{n+nn}{numpy} \PY{k+kn}{as} \PY{n+nn}{np}

\PY{n+nd}{@waLBerla.callback}\PY{p}{(} \PY{l+s}{\PYZdq{}}\PY{l+s}{at\PYZus{}end\PYZus{}of\PYZus{}timestep}\PY{l+s}{\PYZdq{}} \PY{p}{)}
\PY{k}{def} \PY{n+nf}{evaluation}\PY{p}{(}\PY{n}{blockstorage}\PY{p}{,} \PY{n}{bubbles}\PY{p}{)}\PY{p}{:}  
  \PY{c}{\PYZsh{} Distributed evaluation}
  \PY{n}{x\PYZus{}vel\PYZus{}max} \PY{o}{=} \PY{l+m+mi}{0}
  \PY{k}{for} \PY{n}{block} \PY{o+ow}{in} \PY{n}{blockstorage}\PY{p}{:}
    \PY{n}{vel\PYZus{}field} \PY{o}{=} \PY{n}{np}\PY{o}{.}\PY{n}{asarray}\PY{p}{(} \PY{n}{block}\PY{p}{[}\PY{l+s}{\PYZsq{}}\PY{l+s}{velocity}\PY{l+s}{\PYZsq{}}\PY{p}{]} \PY{p}{)}
    \PY{n}{x\PYZus{}vel\PYZus{}max} \PY{o}{=} \PY{n+nb}{max}\PY{p}{(}\PY{n}{vel\PYZus{}field}\PY{p}{[}\PY{p}{:}\PY{p}{,}\PY{p}{:}\PY{p}{,}\PY{p}{:}\PY{p}{,}\PY{l+m+mi}{0}\PY{p}{]}\PY{o}{.}\PY{n}{max}\PY{p}{(}\PY{p}{)}\PY{p}{,} \PY{n}{x\PYZus{}vel\PYZus{}max}\PY{p}{)}
      
  \PY{n}{x\PYZus{}vel\PYZus{}max} \PY{o}{=} \PY{n}{mpi}\PY{o}{.}\PY{n}{reduce}\PY{p}{(} \PY{n}{x\PYZus{}vel\PYZus{}max}\PY{p}{,} \PY{n}{mpi}\PY{o}{.}\PY{n}{MAX} \PY{p}{)}
  \PY{k}{if} \PY{n}{x\PYZus{}vel\PYZus{}max}\PY{p}{:} \PY{c}{\PYZsh{}valid on root only}
    \PY{n}{log}\PY{o}{.}\PY{n}{result}\PY{p}{(}\PY{l+s}{\PYZdq{}}\PY{l+s}{Max X Vel}\PY{l+s}{\PYZdq{}}\PY{p}{,} \PY{n}{x\PYZus{}vel\PYZus{}max}\PY{p}{)}
   
  \PY{c}{\PYZsh{} Gather and evaluate locally}
  \PY{n}{size}\PY{o}{=}\PY{n}{blockstorage}\PY{o}{.}\PY{n}{numberOfCells}\PY{p}{(}\PY{p}{)} 
  \PY{n}{vel\PYZus{}profile\PYZus{}z}\PY{o}{=}\PY{n}{gather\PYZus{}slice}\PY{p}{(} \PY{n}{x}\PY{o}{=}\PY{n}{size}\PY{p}{[}\PY{l+m+mi}{0}\PY{p}{]}\PY{o}{/}\PY{l+m+mi}{2}\PY{p}{,} \PY{n}{y}\PY{o}{=}\PY{n}{size}\PY{p}{[}\PY{l+m+mi}{1}\PY{p}{]}\PY{o}{/}\PY{l+m+mi}{2}\PY{p}{,} \PY{n}{coarsen}\PY{o}{=}\PY{l+m+mi}{4} \PY{p}{)}
  \PY{k}{if} \PY{n}{vel\PYZus{}profile\PYZus{}z}\PY{p}{:} \PY{c}{\PYZsh{}valid on root only}
    \PY{n}{eval\PYZus{}vel\PYZus{}profile}\PY{p}{(}\PY{n}{vel\PYZus{}profile\PYZus{}z}\PY{p}{)}   
\end{Verbatim}

 \mylistingscaption{Simulation evaluation }
 \mylistingslabel{lst:simulation_eval_py}
\end{listing}

Due to the distributed nature of the data, this evaluation step is still somewhat complex. However, we can simplify it in some cases, especially when working with smaller subsets of the data.
Let us for example consider the evaluation of the flow profile in the channel scenario. The velocity profile (as depicted in Figure \ref{fig:scenario_setup}) in the middle
of the channel along a line orthogonal to the flow direction is analyzed. 
From this information we can obtain, for example, the height of the thin liquid layer at the bottom.
To simplify the evaluation routine, we first collect this one dimensional dataset on a single process, which is possible since the 1D 
slice is much smaller than the complete field. In case it is still too big, the slice can also be coarsened, meaning that only every $n$'th cell is gathered.
Then all required data is stored on a single process, enabling a simple serial evaluation of the results.

The same 1D-collection technique is used to determine geometric properties of the lamella in the impining jets scenario.
Since the lamella is placed in the middle of the domain, three 1D line captures along the axes through the domain midpoint are sufficient to determine height, width, and thickness of the lamella.

\subsection{Simulation Control and Steering}

The Python interface can not only be used for domain setup and evaluation but also to interact with the simulation
while it is running using a Python console. This is especially useful during the development process. One can
visualize and analyze the simulation state with plotting libraries (e.g. matplotlib \cite{hunter2007matplotlib}) and then
modify the simulation state interactively.

In case of a serial program, the Python C-API offers high level functions to start an interactive interpreter loop.
For parallel simulations, this approach is not feasible, since every process would start its own console. 
Instead, a custom solution was developed, where one designated process runs the interpreter loop, broadcasting the 
entered commands to all other processes, which are then executed simultaneously. 
The custom interpreter loop reads the user input line by line, until a full command was entered.
This advanced detection has to be done, since Python commands can span multiple lines. After a full command
was detected, it is sent to all other processes using an MPI broadcast operation.

There are two ways to start an interactive console while a simulation is running: In UNIX environments, the user can send a POSIX signal to 
interrupt the simulation. The simulation continues until the end of the current time step, such that all internal data structures are in a consistent
state. Then the Python console is run, using the standard input/output streams. A second method based on TCP sockets can be used when sending POSIX signals 
is not feasible, for example in Windows environments or when starting a parallel simulation using a job scheduler. 
In this case, the program listens for TCP connections. When a client connects, the simulation is interrupted after the current time step such that the user can interact
with the program using a telnet client.

Besides the ability to modify the simulation state interactively, the modification of parameters can of course also be done automatically.
Being able to evaluate the simulation state in Python, we have information available {\em during} the simulation which previously were acquired in a post-processing step.
This information can be used to modify parameters while the simulation is running, effectively implementing a feedback loop. 
In the channel flow scenario, this can be used to investigate foam stability when the inflow boundary is switched on or off. Evaluation routines
are used to determine if the simulation has reached a steady state, then the driving pressure boundary can be modified.

\subsection{Summary}

To summarize, the Python interface has greatly simplified the entire simulation toolchain.

\begin{figure}
 \begin{center}
    \subfigure[]{
      \resizebox*{5cm}{!}{  \includegraphics{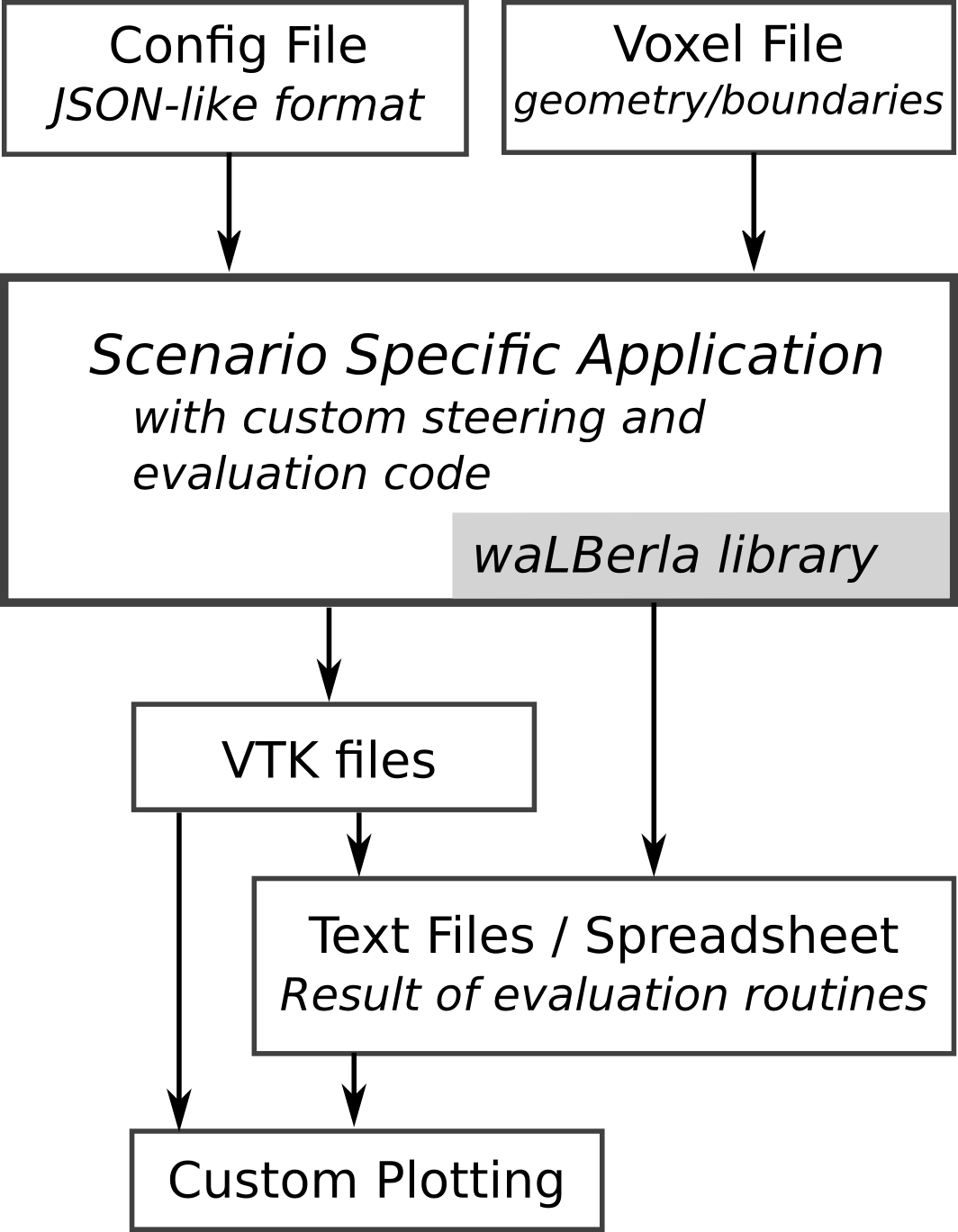}  } 
     }
     \hspace{0.8cm}
    \subfigure[]{
      \resizebox*{6cm}{!}{  \includegraphics{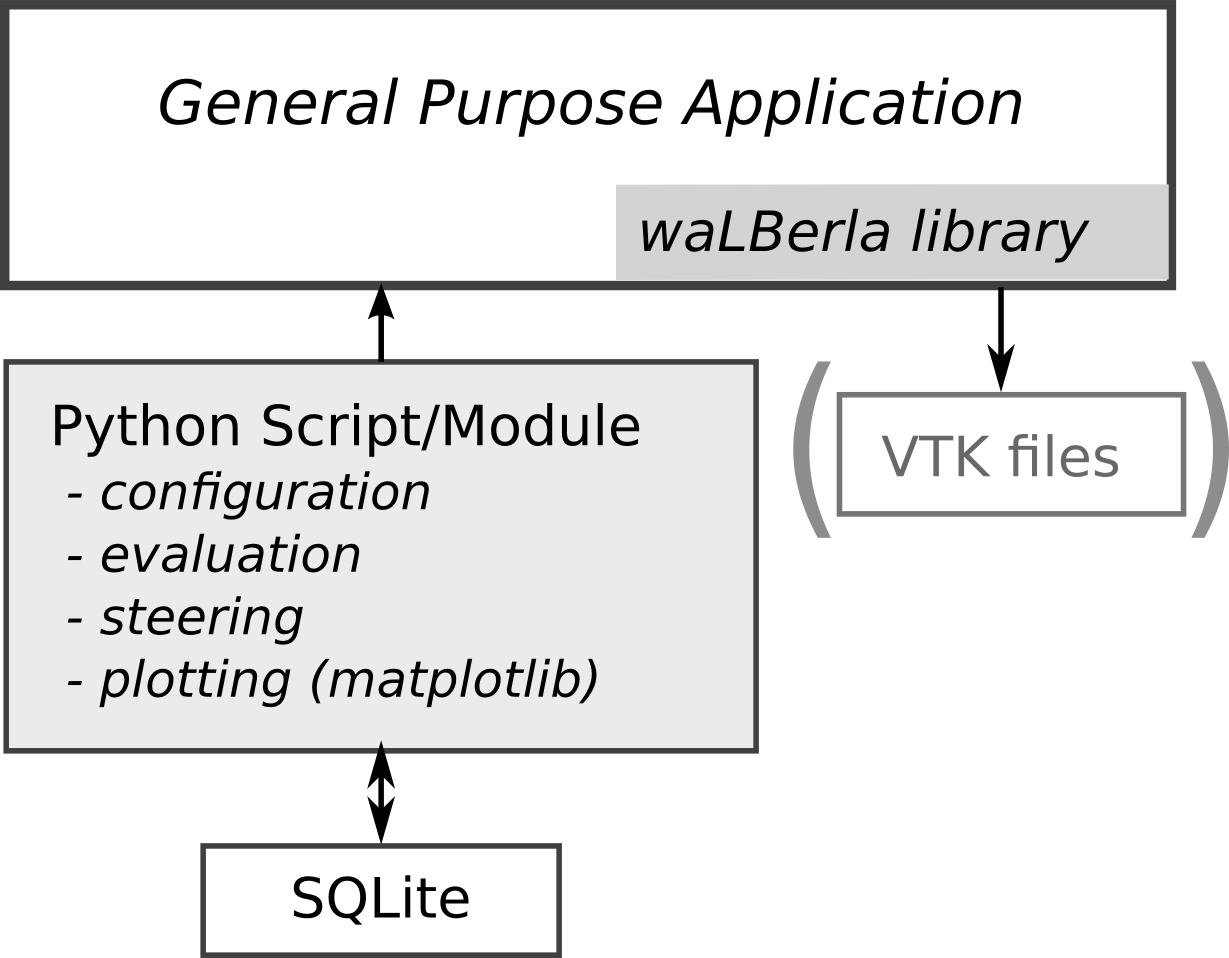} 
     }
    }
     \caption{ Typical simulation workflow without usage (left) and with usage (right) of Python Interface}
     \label{fig:toolchain_python}
 \end{center}
\end{figure}

A schematic of a typical workflow of using \Walberla{} without the Python extension is depicted in Figure \ref{fig:toolchain_python}a.
The configuration is supplied in two different files: in one voxel file for specifying domain geometry and boundary information and a second text based
parameter file defining options as a hierarchical set of key-value pairs.

The C++ simulation code itself is also tailored to the scenario, including custom steering and evaluation functions.
Changing the scenario involves recompilation of the binary.
It is not possible to write a general purpose application in this case since the customization options of the configuration file are usually not sufficient.
To analyze the simulation, either the complete flow field is written out in a VTK file for later post-processing, or results are stored in custom text files or
spreadsheets.

Figure \ref{fig:toolchain_python}b, in contrast, shows how this workflow is simplified using the Python scripting capabilities of \Walberla{}.
Whereas previously the description of how to set up and evaluate one scenario was spread out over many files, all this information is now located 
in a single Python file, or in complex cases, in a Python module. 
Results can already be evaluated during simulation runs with less development effort compared to C++. The extracted quantities of interest
are stored in a relational database, typically using SQLite due to its low configuration overhead. Of course, the complete simulation data can still
be written out to VTK files, but in many cases this is not necessary.
Visualization and plotting of the collected results can be implemented in the same script, leading to a compact and reusable collection of all
information related to a specific simulation setup.


\section{Conclusion}

We showed the advantages of coupling the \Walberla{} C++ framework to Python, implementing performance critical parts in C++
and higher level functionality, like domain setup, simulation control, and evaluation of results in Python.
We simplified and automated the simulation workflow, starting from scenario definition up to plotting of the results.
The flexibility and expressive power of Python enables the user to develop code faster compared to C++. It is therefore also suitable for
prototyping of new methods or boundary conditions, a task that previously was done using tools like {\em Matlab}.
The Python interface of \Walberla{} makes the framework more attractive for domain experts, which typically are not familiar with C++ programming.

\section*{Acknowledgments}
We are grateful to the Leibniz Rechenzentrum in Garching and the J\"ulich Supercomputing Center for providing computational resources.

\bibliographystyle{gPAA}
\bibliography{pyWaLBerla}

\begin{thebibliography}{10}
\providecommand{\url}[1]{\texttt{#1}}
\providecommand{\urlprefix}{URL }

\bibitem{dongarra2007impact}
J. Dongarra, D. Gannon, G. Fox, and K. Kennedy,  \emph{The impact of multicore
  on computational science software}, CTWatch Quarterly 3 (2007), pp. 1--10.

\bibitem{heroux2005overview}
M.A. Heroux, R.A. Bartlett, V.E. Howle, R.J. Hoekstra, J.J. Hu, T.G. Kolda,
  R.B. Lehoucq, K.R. Long, R.P. Pawlowski, E.T. Phipps, \emph{et~al.},
  \emph{An overview of the trilinos project}, ACM Transactions on Mathematical
  Software (TOMS) 31 (2005), pp. 397--423.

\bibitem{PyTrilinos}
M. Sala, W. Spotz, and M. Heroux,  \emph{{PyTrilinos}: High-performance
  distributed-memory solvers for {Python}}, ACM Transactions on Mathematical
  Software (TOMS) 34 (2008).

\bibitem{ketcheson2012pyclaw}
D.I. Ketcheson, K. Mandli, A.J. Ahmadia, A. Alghamdi, M.Q. Lunade~, M. Parsani,
  M.G. Knepley, and M. Emmett,  \emph{Pyclaw: Accessible, extensible, scalable
  tools for wave propagation problems}, SIAM Journal on Scientific Computing 34
  (2012), pp. C210--C231.

\bibitem{alghamdi2011petclaw}
A. Alghamdi, A. Ahmadia, D.I. Ketcheson, M.G. Knepley, K.T. Mandli, and L.
  Dalcin, \emph{PetClaw: A scalable parallel nonlinear wave propagation solver
  for Python}, in \emph{Proceedings of the 19th High Performance Computing
  Symposia}, 2011, pp. 96--103.

\bibitem{mandli2011using}
K. Mandli, A. Alghamdi, A. Ahmadia, D.I. Ketcheson, and W. Scullin, \emph{Using
  python to construct a scalable parallel nonlinear wave solver}, in
  \emph{Proceedings of the 10th Python in science Conf.(SciPy 2011)}, 2011.

\bibitem{palabosWebsite}
 Palabos, Accessed: 2015-11-01, \url{http://www.palabos.org/}.

\bibitem{januszewski2014sailfish}
M. Januszewski and M. Kostur,  \emph{Sailfish: a flexible multi-gpu
  implementation of the lattice boltzmann method}, Computer Physics
  Communications 185 (2014), pp. 2350--2368.

\bibitem{terrel2011pythonCodeGeneration}
A.R. Terrel,  \emph{From equations to code: Automated scientific computing},
  Computing in Science \& Engineering 13 (2011), pp. 78--82.

\bibitem{logg2012fenics}
A. Logg, K.A. Mardal, and G. Wells,  \emph{Automated solution of differential
  equations by the finite element method: The FEniCS book}, Vol.~84, Springer
  Science \& Business Media, 2012.

\bibitem{kirby2006compiler}
R.C. Kirby and A. Logg,  \emph{A compiler for variational forms}, ACM
  Transactions on Mathematical Software (TOMS) 32 (2006), pp. 417--444.

\bibitem{Feichtinger2011105}
C. Feichtinger, S. Donath, H. K\"ostler, J. G\"otz, and U. R\"ude,
  \emph{{WaLBerla: HPC software design for computational engineering
  simulations}}, Journal of Computational Science 2 (2011), pp. 105 -- 112,
  \urlprefix\url{http://www.sciencedirect.com/science/article/pii/S1877750311000111}.

\bibitem{walberlaWebsite}
 {waLBerla Framework},  \emph{http://walberla.net}  (2015).

\bibitem{Aidun2010}
C.K. Aidun and J.R. Clausen,  \emph{Lattice-{Boltzmann} method for complex
  flows}, Annual Review of Fluid Mechanics 42 (2010), pp. 439--472,
  \urlprefix\url{http://www.annualreviews.org/doi/pdf/10.1146/annurev-fluid-121108-145519}.

\bibitem{Qian1992}
Y. Qian, D. d'Humieres, and P. Lallemand,  \emph{Lattice {BGK} models for
  navier-stokes equation}, EPL (Europhysics Letters) 17 (1992), p. 479,
  \urlprefix\url{http://iopscience.iop.org/0295-5075/17/6/001}.

\bibitem{Ginzburg2008}
I. Ginzburg, F. Verhaeghe, and D. d'Humieres,  \emph{Two-relaxation-time
  lattice {Boltzmann} scheme: About parametrization, velocity, pressure and
  mixed boundary conditions}, Communications in computational physics 3 (2008),
  pp. 427--478,
  \urlprefix\url{https://lirias.kuleuven.be/handle/123456789/218788}.

\bibitem{Ginzburg2008a}
I. Ginzburg, F. Verhaeghe, and D. d'Humieres,  \emph{Study of simple
  hydrodynamic solutions with the two-relaxation-times lattice {Boltzmann}
  scheme}, Communications in computational physics 3 (2008), pp. 519--581,
  \urlprefix\url{https://lirias.kuleuven.be/handle/123456789/218787}.

\bibitem{korner2005lattice}
C. K{\"o}rner, M. Thies, T. Hofmann, N. Th{\"u}rey, and U. R{\"u}de,
  \emph{Lattice boltzmann model for free surface flow for modeling foaming},
  Journal of Statistical Physics 121 (2005), pp. 179--196.

\bibitem{donath2009localized}
S. Donath, C. Feichtinger, T. Pohl, J. G{\"o}tz, and U. R{\"u}de,
  \emph{Localized parallel algorithm for bubble coalescence in free surface
  lattice-boltzmann method}, in \emph{Euro-Par 2009 Parallel Processing},
  Springer, 2009, pp. 735--746.

\bibitem{Goetz2008}
J. G\"otz, C. Feichtinger, K. Iglberger, S. Donath, and U. R\"ude, \emph{{Large
  scale simulation of fluid structure interaction using Lattice Boltzmann
  methods and the `physics engine'}}, in \emph{Proceedings of the 14th Biennial
  Computational Techniques and Applications Conference, CTAC-2008}, ANZIAM J.,
  Vol.~50, Oct., \url
  {http://anziamj.austms.org.au/ojs/index.php/ANZIAMJ/article/view/1445}
  [October 29, 2008], 2008, pp. C166--C188.

\bibitem{markl2015numerical}
M. Markl, R. Ammer, U. R{\"u}de, and C. K{\"o}rner,  \emph{Numerical
  investigations on hatching process strategies for powder-bed-based additive
  manufacturing using an electron beam}, The International Journal of Advanced
  Manufacturing Technology 78 (2015), pp. 239--247.

\bibitem{schornbaum2015massively}
F. Schornbaum and U. R{\"u}de,  \emph{Massively parallel algorithms for the
  lattice boltzmann method on non-uniform grids}, arXiv preprint
  arXiv:1508.07982 subbmited to SIAM Journal on Scientific Computing  (2015).

\bibitem{Burstedde2014forestclaw}
C. Burstedde, D. Calhoun, K. Mandli, and A.R. Terrel,  \emph{{ForestClaw}:
  {Hybrid} forest-of-octrees {AMR} for hyperbolic conservation laws}, Advances
  in Parallel Computing 25 (2014), pp. 253 -- 262.

\bibitem{Goetz:2010:ParComp}
J. G\"otz, K. Iglberger, C. Feichtinger, S. Donath, and U. R\"ude,
  \emph{{Coupling multibody dynamics and computational fluid dynamics on 8192
  processor cores}}, Parallel Computing 36 (2010), pp. 142--141.

\bibitem{feichtinger2015performance}
C. Feichtinger, J. Habich, H. K{\"o}stler, U. R{\"u}de, and T. Aoki,
  \emph{Performance modeling and analysis of heterogeneous lattice boltzmann
  simulations on cpu--gpu clusters}, Parallel Computing 46 (2015), pp. 1--13.

\bibitem{godenschwager2013}
C. Godenschwager, F. Schornbaum, M. Bauer, H. K{\"o}stler, and U. R{\"u}de,
  \emph{{A framework for hybrid parallel flow simulations with a trillion cells
  in complex geometries}}, in \emph{{Proceedings of SC13: International
  Conference for High Performance Computing, Networking, Storage and
  Analysis}}, 2013, p.~35.

\bibitem{LSS:Bauer:2015}
M. Bauer, J. Hötzer, P. Steinmetz, M. Jainta, M. Berghoff, F. Schornbaum, C.
  Godenschwager, H. Köstler, B. Nestler, and U. Rüde,  \emph{Massively
  parallel phase-field simulations for ternary eutectic directional
  solidification}, arXiv preprint, accepted at Supercomputing 2015  (2015),
  \urlprefix\url{http://arxiv.org/abs/1506.01684}.

\bibitem{boostWebsite}
 {boost C++ libraries},  \emph{http://www.boost.org/}  (2015).

\bibitem{boostPythonWebsite}
 {boost.python library},
  \emph{http://www.boost.org/doc/libs/1\_59\_0/libs/python/doc/}  (2015).

\bibitem{van2011numpy}
S. Van Der~Walt, S.C. Colbert, and G. Varoquaux,  \emph{The numpy array: a
  structure for efficient numerical computation}, Computing in Science \&
  Engineering 13 (2011), pp. 22--30.

\bibitem{pythonBufferProtocol}
 {Python Buffer Protocol},  \emph{https://docs.python.org/3/c-api/buffer.html}
  (2015).

\bibitem{junk2006}
M. Junk and D. Kehrwald,  \emph{On the relation between lattice variables and
  physical quantities in lattice boltzmann simulations}, ITWM Report  (2006).

\bibitem{pintWebsite}
 {Python Units Library Pint},  \emph{http://pint.readthedocs.org/}  (2015).

\bibitem{jones2001scipy}
E. Jones, T. Oliphant, and P. Peterson,  \emph{Scipy: Open source scientific
  tools for python}, http://www. scipy. org/  (2001).

\bibitem{joyner2012sympy}
D. Joyner, O. {\v{C}}ert{\'\i}k, A. Meurer, and B.E. Granger,  \emph{Open
  source computer algebra systems: Sympy}, ACM Communications in Computer
  Algebra 45 (2012), pp. 225--234.

\bibitem{henderson2004paraview}
A. Henderson, J. Ahrens, and C. Law,  \emph{The ParaView Guide}, Kitware
  Clifton Park, NY, 2004.

\bibitem{hunter2007matplotlib}
J.D. Hunter,  \emph{Matplotlib: A 2d graphics environment}, Computing in
  Science \& Engineering 9 (2007), pp. 0090--95.

\end{thebibliography}

\end{document}